\pdfoutput=1
\documentclass[5p,times]{elsarticle}

\usepackage{lineno,hyperref}
\usepackage{bm}
\usepackage{amsmath}
\usepackage{color}
\usepackage{algorithm}
\usepackage{algorithmic}
\usepackage{multirow}
\usepackage{hhline}
\usepackage{makecell}
\usepackage{tikz}
\modulolinenumbers[5]

\journal{Computers \& Fluids}

\begin{document}

\begin{frontmatter}

\title{Wall modeling via function enrichment: extension to detached-eddy simulation}

\author[]{Benjamin Krank}
\ead{krank@lnm.mw.tum.de}
\author[]{Martin Kronbichler}
\ead{kronbichler@lnm.mw.tum.de}
\author[]{Wolfgang A. Wall\corref{correspondingauthor1}}
\cortext[correspondingauthor1]{Corresponding author at: Institute for Computational Mechanics, Technical University of Munich, Boltzmannstr. 15, 85748 Garching, Germany. Tel.: +49 89 28915300; fax: +49 89 28915301}
\ead{wall@lnm.mw.tum.de}
\address{Institute for Computational Mechanics, Technical University of Munich,\\ Boltzmannstr. 15, 85748 Garching, Germany}

\begin{abstract}
We extend the approach of wall modeling via function enrichment to detached-eddy simulation. The wall model aims at using coarse cells in the near-wall region by modeling the velocity profile in the viscous sublayer and log-layer. However, unlike other wall models, the full Navier--Stokes equations are still discretely fulfilled, including the pressure gradient and convective term. This is achieved by enriching the elements of the high-order discontinuous Galerkin method with the law-of-the-wall. As a result, the Galerkin method can ``choose'' the optimal solution among the polynomial and enrichment shape functions. The detached-eddy simulation methodology provides a suitable turbulence model for the coarse near-wall cells. The approach is applied to wall-modeled LES of turbulent channel flow in a wide range of Reynolds numbers. Flow over periodic hills shows the superiority compared to an equilibrium wall model under separated flow conditions.
\end{abstract}


\end{frontmatter}

\section{Introduction}
Wall modeling via function enrichment is a spatial discretization technique that allows the resolution of the sharp boundary layer gradients present in high-Reynolds-number flows with relatively coarse meshes. The basic idea is to make use of the flexibility in Galerkin methods regarding the choice of the solution space: a few additional, problem-tailored shape functions are used to approximate the solution, in addition to the common polynomials. Using these enriched elements, the full Navier--Stokes equations are solved in the whole boundary layer in a consistent manner. As a result, the wall model can take into account high adverse pressure gradients and convective effects, unlike most other wall modeling approaches.

The idea of wall modeling via function enrichment was proposed by Krank and Wall~\cite{Krank16} within the continuous Galerkin method (standard FEM) as a wall modeling technique for large-eddy simulation (LES). While that work showed promising results in separated flows, the limiting factor in terms of accuracy was the turbulence model employed in the near-wall region. A residual-based approach was used, supported by a structural LES model in the outer layer, a model that was originally not intended for underresolved boundary-layer simulations.
The wall modeling approach was since applied in conjunction with RANS~\cite{Krank16c} employing the Spalart--Allmaras (SA) model within the high-order discontinuous Galerkin (DG) method.

In this article, we show that the widely used delayed detached-eddy simulation (DDES) methodology~\cite{Spalart06} may be used to model the unresolved turbulence in the near-wall region in wall modeling via function enrichment. This can be done by extending the implementation of the SA model~\cite{Krank16c} in a straightforward way. The idea of the original DES approach~\cite{Spalart97} is that the wall distance function $y$ present in the SA model is limited with a characteristic cell length $\Delta$ according to
\begin{equation}
y_{\mathrm{DES}} = \min(y,C_{\mathrm{DES}}\Delta),
\end{equation}
where the parameter $C_{\mathrm{DES}}$ has been calibrated to $C_{\mathrm{DES}}=0.65$ and the grid length scale is defined as the maximum of the cell length over the space dimensions $\Delta=\max(\Delta_x,\Delta_y,\Delta_z)$~\cite{Shur99}. As a result, the RANS model acts as a one-equation LES subgrid model if $y>C_{\mathrm{DES}} \Delta$. DDES represents an enhancement of that methodology by defining the wall-distance parameter as
\begin{equation}
y_{\mathrm{DDES}} = y - f_d \max(0,y-C_{\mathrm{DES}}\Delta),
\end{equation}
with the functions
\begin{align}
f_d&=1-\mathrm{tanh}\left((8r_d)^3\right),\\
r_d&=\frac{\nu+\nu_t}{\sqrt{(\nabla \bm{u})_{ij}(\nabla \bm{u})_{ij}} \kappa^2 y^2},
\end{align}
where $\bm{u}$ is the velocity vector, $\nu$ the kinematic and $\nu_t$ the eddy viscosity, and $\kappa=0.41$.
(D)DES is widely used in research and industry, see, e.g.,~\cite{Frohlich08,Piomelli08} and is today even used for the aerodynamics of entire vehicles~\cite{Blacha16} due to its good accuracy in separated flows and the ability to investigate acoustic noise sources in the flow. Regarding the application of DES, two main branches are frequently used. The original idea was to simulate the whole boundary layer in RANS mode and to compute free shear layers in LES mode only~\cite{Spalart97}. As an alternative, DES can be seen as an approach to wall-modeled LES (WMLES), in which only the inner layer is computed in RANS mode and the outer boundary layer in LES mode, see, e.g.,~\cite{Nikitin00}.

Wall modeling via function enrichment has the potential of significantly reducing the computational cost of (D)DES. The grid saving of the standard (D)DES in comparison to LES is achieved by using relatively coarse meshes in the wall-parallel directions of up to $0.1\delta$ (WMLES) and $\delta$ (classical DES)~\cite{Larsson16} with the boundary layer thickness $\delta$. The wall-normal direction necessitates many grid points in order to resolve the laminar sublayer due to the requirement of placing the first off-wall node at $y^+_1\sim 1$, however. For example, if a boundary layer of a thickness of $10{,}000$ wall units is computed with a constant grid stretching factor of 1.15~\cite{Nikitin00}, a total of 53 grid layers would be required. This is a quite high cost compared to the relatively low engineering interest in that region. Wall modeling via function enrichment allows the first grid point to be located in the range $y^+_1\sim 10$ to $100$, saving 17--33 grid layers for that example, without noteworthy loss in accuracy, in addition to much better conditioned equation systems through the lower grid anisotropy.

In the next section, we give details on how the enrichment shape functions are constructed. In Section~\ref{sec:num}, the high-order DG code employed for the validation is outlined and numerical examples are presented in Section~\ref{sec:examples}.

\section{Wall modeling via function enrichment}
The primary idea of wall modeling via function enrichment is as follows. In a single element row at the wall, the discrete velocity solution $\bm{u}_h$ is composed of two parts, the standard polynomial component, $\bar{\bm{u}}_h$, and an additional enrichment component, $\widetilde{\bm{u}}_h$, yielding
\begin{equation}
\bm{u}_h(\bm{x},t)=\bar{\bm{u}}_h(\bm{x},t)+\widetilde{\bm{u}}_h(\bm{x},t).
\label{eq:dg_xw_rans:space}
\end{equation}
The polynomial component is given in each cell as an FE-expansion according to
\begin{equation}
\bar{\bm{u}}_h(\bm{x},t)=\sum_{B \in N^{k}} N_B^{k}(\bm{x}) \bar{\bm{u}}_B(t)
\label{eq:dg_xw_rans:std_fe}
\end{equation}
with the shape functions $N_B^{k}$ of polynomial degree $k$ and corresponding degrees of freedom $\bar{\bm{u}}_B$. There are several ways of constructing the enrichment component. In its simplest form, an enrichment function $\psi$ is weighted in each element with one additional node $\widetilde{\bm{u}}_{0}$, i.e., one degree of freedom per space dimension, 
with
\begin{equation}
\widetilde{\bm{u}}_h(\bm{x},t)=\psi(\bm{x},t)
\widetilde{\bm{u}}_{0}(t).
\label{eq:dg_xw_rans:enrichment}
\end{equation}
The enrichment function can additionally be weighted using a low-order polynomial to yield a higher level of flexibility in the function space~\cite{Krank16,Krank16c}, which is not considered herein.
It is this enrichment function that is responsible for the efficiency of the approach. By taking $\psi$ as a wall function, the solution space of the Galerkin method is capable of resolving a sharp attached boundary layer with very few degrees of freedom. It is noted that this wall function is not prescribed as a solution, but the Galerkin method automatically ``chooses'' the best possible solution within the high-order polynomials and the enrichment component in a least squares sense. As a wall function, we consider Spalding's law~\cite{spalding61} in the form
\begin{equation}
y^+=\frac{\psi}{\kappa}+e^{-\kappa B}\left(e^{\psi}-1-\psi-\frac{\psi^2}{2!}-\frac{\psi^3}{3!}-\frac{\psi^4}{4!}\right),
\label{eq:dg_xw_rans:spald}
\end{equation}
with $\kappa=0.41$ and $B=5.17$, as it was implemented in~\cite{Krank16}. Several alternative wall functions have been discussed in~\cite{Krank16c}. In the wall-normal direction, Spalding's law scales with the wall coordinate $y^+=y u_{\tau}/ \nu$ with $u_{\tau}=\sqrt{\tau_w/\rho}$ and the density $\rho$ such that the wall shear stress $\tau_w$ is represented correctly. In turn, the wall function has to be adapted according to the local wall shear stress in the numerical method, and its temporal evolution has to be taken into account.

We have developed an algorithm in~\cite{Krank16,Krank16c}, which enables such an adaptation. Therein, the wall shear stress is computed on discrete nodes via the velocity derivative according to
\begin{equation}
\tau_{w,B}=\frac{\lVert\int_{\partial \Omega^D} N_B^{c,m}(\bm{x})\rho \nu \frac{\partial \bm{u}_h}{\partial y} \big|_{y=0} \ \mathrm{d}A \rVert}{\int_{\partial \Omega^D} N_B^{c,m}(\bm{x}) \ \mathrm{d}A},
\end{equation}
with linear continuous shape functions $N_B^{c,m}$ of degree $m=1$. The nodal values are interpolated by
\begin{equation}
\tau_{w,h}=\sum_{B \in N^{c,m}}N_B^{c,m} \tau_{w,B},
\end{equation}
yielding a continuous representation of the wall shear stress. Through the choice of $m=1$, the wall shear stress $\tau_{w,h}$ is a coarsened field, since usually $k>1$; the coarsening is mandatory because the wall functions are relations for the mean quantities, meaning that the mean wall shear stress is related to the mean velocity, and the average wall shear stress would otherwise be overpredicted, see Reference~\cite{Krank16}. This field is updated prior to each time step, such that the function space of the velocity changes continuously and adapts to the local flow conditions. Further details on the adaptation algorithm are given in~\cite{Krank16c}. Near separation or reattachment locations, it may happen that the wall shear stress becomes zero, which renders the function space linear dependent. However, considering that the first off-wall point is located very close to the wall in terms of $y^+$ at these locations, a simple and consistent solution is to temporally ``switch off'' the enrichment in the respective cells, see~\cite{Krank16c} for details and~\cite{Krank17} for an evaluation of the method in WMLES in the context of another turbulence modeling approach. If wall shear stress becomes larger at these locations at a later instance, the enrichment is ``switched on'' again.

Finally, we comment on the two additional variables, which have to be discretized: the pressure and the working variable of the SA model. Both variables do not exhibit high gradients at the wall, such that they are represented sufficiently well by the standard FE space only, according to~\cite{Krank16c}.

\section{Numerical method}
\label{sec:num}
The present wall modeling approach may be implemented in any FEM and DG flow solver. In this work, we consider the implementation of the incompressible Navier--Stokes equations with the SA model in~\cite{Krank16c}, which in turn is based on the incompressible high-performance high-order semi-explicit DG code INDEXA~\cite{Krank16b}. An extension of the present wall modeling approach to the compressible Navier--Stokes equations would be straightforward, since high gradients are commonly not present in the energy variable, such that the latter may be considered analogous to the pressure variable herein. Numerical methods based on the continuous FEM would require a small modification of the enrichment component as described in~\cite{Krank16}.

The solver is based on weak forms, which are described in detail in~\cite{Krank16c}. These weak forms include volume and surface terms, that have to be integrated over cells and faces. The integrals are in our solver evaluated using the high-performance kernels by Kronbichler and Kormann~\cite{Kronbichler12} within the deal.II finite element library~\cite{Arndt17}. In particular, the integrals have polynomilal and nonpolynomial paths, the latter due to the nonpolynomial character of the enrichment function. The polynomial paths are integrated using the quadrature formulas given in~\cite{Krank16b} and are evaluated exactly on affine cells. The nonpolynomial contributions have to be evaluated with more quadrature points, in particular in the wall-normal direction~\cite{Krank16}. From our extensive experience with wall modeling via function enrichment, we can give the following guide lines: If the enriched cells extend up to approximately $y_{1e}^+=90$ in the statistical quantities, 8 quadrature points in the wall-normal direction are sufficient. Further we have $y_{1e}^+<110$ (10 points), $y_{1e}^+=130$ (12 points), $y_{1e}^+=200$ (17 points); see also the monograph~\cite{Krank18} for further details. All simulation cases presented herein use an adaptive time stepping method presented in~\cite{Krank16c} with a temporal accuracy of second order, a Courant number of $\mathrm{Cr}=0.14$, and a diffusion number of $D=0.02$. In the particular formulation used with the enrichment, the solver has a formal spatial order of accuracy of $k$. Finally, we note that we apply no-slip boundary conditions weakly according to~\cite{Krank16b} in all steps of the scheme for the examples presented in this article, which limits the width of the first off-wall cell to a few hundred wall units, as the no-slip condition would otherwise be violated severely.

The increasing resolution power of the DG scheme with increasing polynomial degree should be taken into account in the (D)DES grid length scale $\Delta$~\cite{Wurst13}. Based on the analysis of the resolution power of DG schemes performed in~\cite{Moura16}, we choose
\begin{equation}
\Delta = \frac{\Delta_e}{k+1}
\label{eq:k_scaling}
\end{equation}
as a length scale, based on the respective cell size $\Delta_e$, in contrast to the choice of the factor of $1/k$ chosen in Reference~\cite{Wurst13}.

\begin{table}[t]
\caption{Overview of simulation cases for the turbulent channel flow. The number of polynomial grid points per direction $i$ is $N_i=(k+1)N_{ie}$ with the number of cells per direction $N_{ie}$ and the polynomial degree $k=4$, $\Delta  y_{1e}^+$ is the thickness of the first off-wall cell, in which the enrichment is active, $y = C_{\mathrm{DES}} \Delta$ is the RANS--LES switching location in terms of channel half-height $\delta$, and err($\tau_w$) is the relative error of the computed wall shear stress.}
\label{tab:ch_flows}
\begin{tabular*}{\linewidth}{l @{\extracolsep{\fill}} l l l l l}
\hline
 $Re_{\tau}$   & $N_{1e} {\times} N_{2e} {\times} N_{3e}   $   & $\gamma$ & $\Delta  y_{1e}^+   $  & $y = C_{\mathrm{DES}} \Delta $ & err($\tau_w$)
\\ \hline \noalign{\smallskip}
$395$  & $16 {\times} 8 {\times} 8$    & $0.8$ & $76$ & $0.05\delta$ & $0.4\%$\\
 $950$  & $16 {\times} 8 {\times} 8$    & $1.6$ & $91$& $0.05\delta$ & $4.9\%$ \\
 $2{,}000$  & $16 {\times} 8 {\times} 8$    & $1.9$ & $137$& $0.05\delta$ & $-4.5 \%$\\
            & $16 {\times} 16 {\times} 8$    & $1.9$ & $54$& $0.05\delta$ & $0.8\%$\\
            & $32 {\times} 16 {\times} 16$    & $1.9$ & $54$& $0.025\delta$ & $1.3\%$\\
  $5{,}200$  & $16 {\times} 16 {\times} 8$    & $2.2$ & $93$& $0.05\delta$ & $0.9\%$\\
   $10{,}000$\  & $16 {\times} 16 {\times} 8$    & $2.5$ & $116$& $0.05\delta$ & $-1.9\%$\\
 $20{,}000$ \ & $16 {\times} 24 {\times} 8$    & $2.5$& $139$& $0.05\delta$ & $1.1\%$\\ 
 $50{,}000$ \ & $16 {\times} 40 {\times} 8$     & $2.5$& $191$& $0.05\delta$ & $-1.4\%$\\ 
\noalign{\smallskip}
\hline
\end{tabular*}
\end{table}

\begin{figure}[tb]
\centering
\includegraphics[trim= 10mm 40mm 10mm 40mm,clip,width=0.95\linewidth]{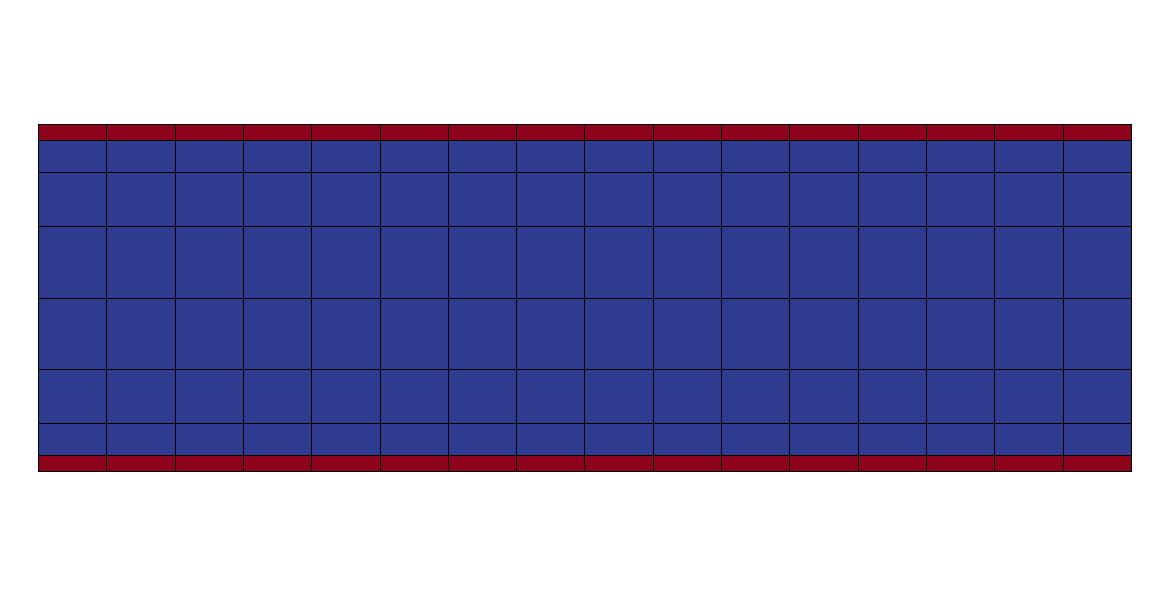}
\caption{Mesh for turbulent channel flow at $Re_{\tau}=950$. Red indicates enriched cells and blue standard polynomial cells, i.e., a single layer of cells at the wall is enriched. In each cell, the solution consists of a polynomial of $4^{\mathrm{th}}$ degree plus one enrichment shape function in the enriched cells.}
\label{fig:ch_mesh}
\includegraphics[trim= 10mm 40mm 10mm 30mm,clip,width=0.95\linewidth]{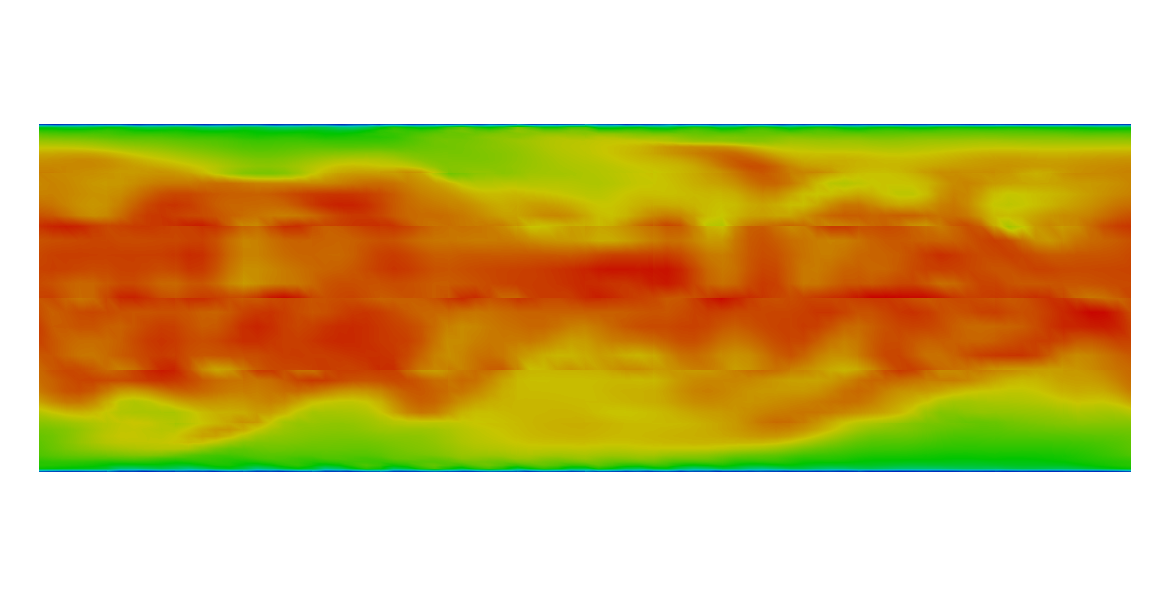}
\caption{Instantaneous numerical solution of turbulent channel flow at $Re_{\tau}=950$ via velocity magnitude. Red indicates high and blue low values.}
\label{fig:ch_flow}
\end{figure}

\section{Numerical examples}
\label{sec:examples}

Wall modeling via function enrichment is assessed by considering DDES in the WMLES branch. In the first example, we investigate the method for attached equilibrium boundary layer flows present in turbulent channel flow. The second example considers flow over periodic hills in order to analyze the behavior of the enrichment in conjunction with DDES in a nonequilibrium flow. As a result of earlier studies~\cite{Krank16b,Krank17b}, the polynomial degree of $k=4$ has proven to be a good compromise between accuracy and time-to-solution, so this polynomial degree is used for all simulation cases presented.

\begin{figure*}[t]
\centering
\begin{minipage}[b]{0.497\linewidth}
\centering
\includegraphics[trim= 8mm 9mm 13mm 10mm,clip,width=1.\textwidth]{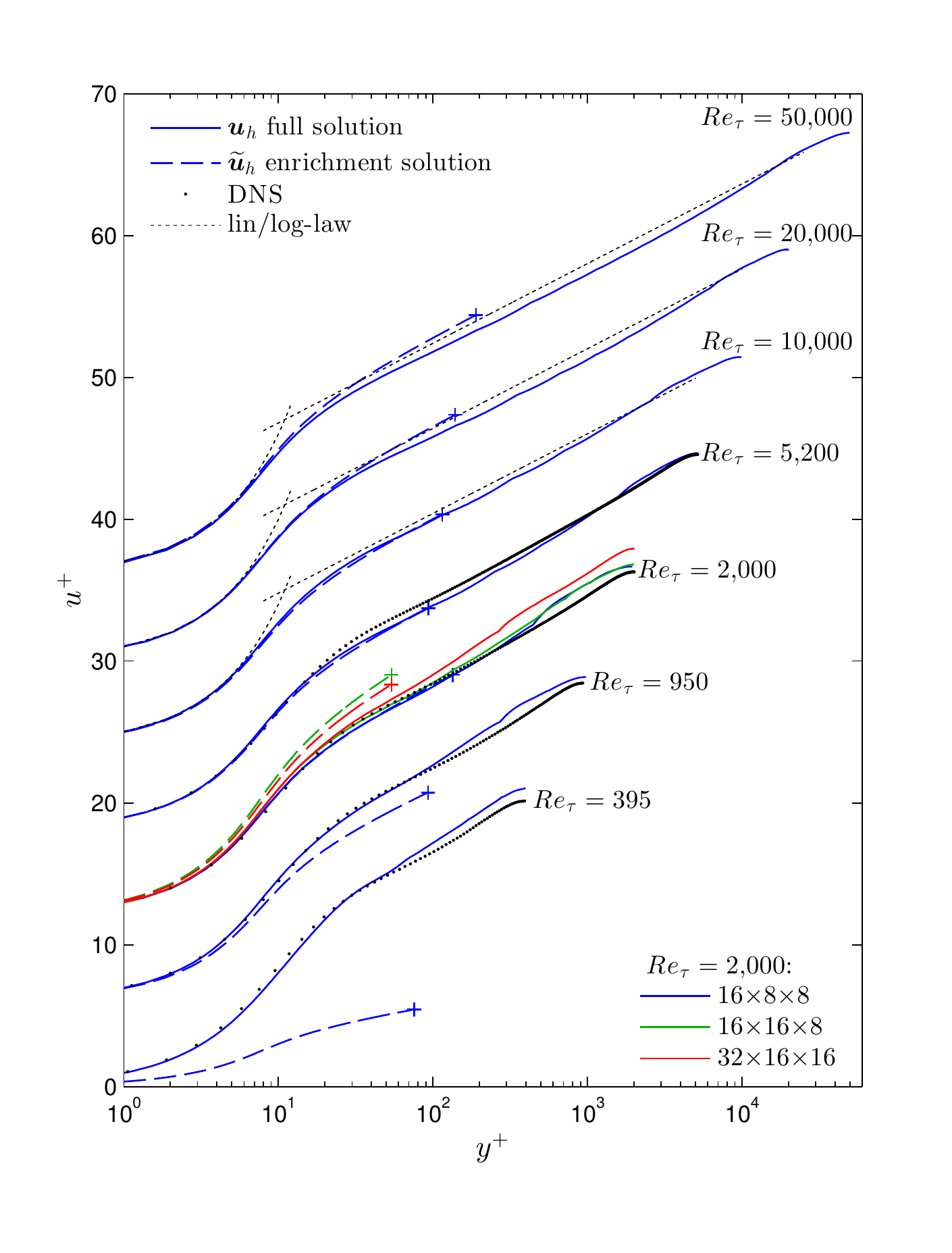}
\end{minipage}
\begin{minipage}[b]{0.497\linewidth}
\centering
\includegraphics[trim= 8mm 9mm 13mm 10mm,clip,width=1.\textwidth]{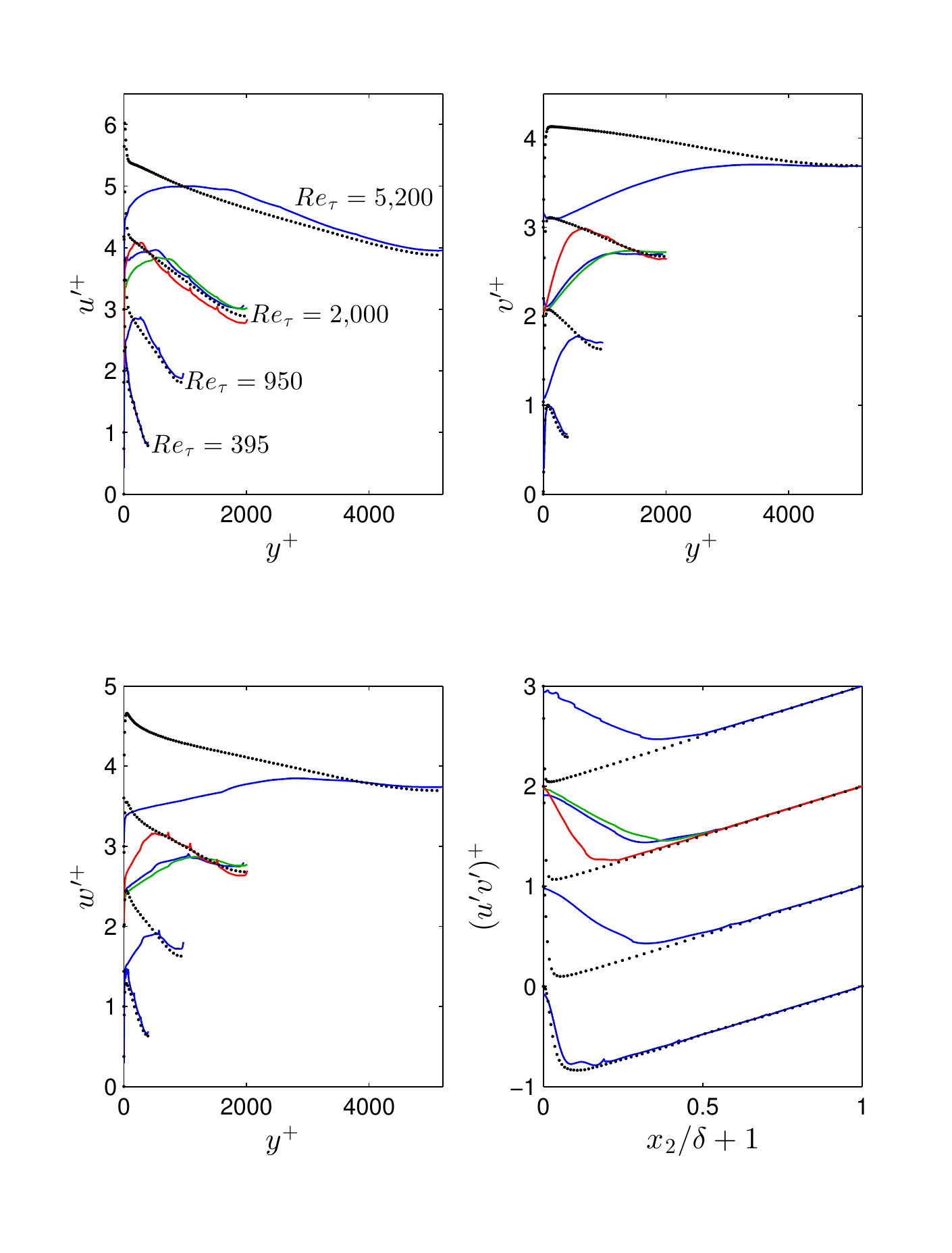}
\end{minipage}
\caption{DDES (WMLES) of turbulent channel flow at several Reynolds numbers. Mean velocity (left) and RMS-velocities as well as Reynolds shear stress (right). All quantities are normalized according to $u^+=\langle u_1\rangle/u_{\tau}$, $u^{\prime+}= \sqrt{\langle u_1^{\prime2}\rangle}/u_{\tau}$, $v^{\prime+}= \sqrt{\langle u_2^{2}\rangle}/u_{\tau}$, $w^{\prime+}=\sqrt{\langle u_3^{2}\rangle}/u_{\tau}$, and $(u^{\prime}v^{\prime})^{+}=\langle u_1 u_2\rangle/u_{\tau}^2$.}
\label{fig:ch_u}
\end{figure*}

\subsection{Turbulent channel flow}
We consider flow in a stream- and spanwise periodic channel of the dimensions $2\pi\delta{\times}2\delta{\times}\pi\delta$ in streamwise, wall-normal, and spanwise direction, respectively, with the channel half-height $\delta$. The flow is driven by a constant body force, which is derived from the nominal quantities. We investigate this flow in a wide range of friction Reynolds numbers $Re_{\tau}=u_{\tau}\delta/\nu$, which are chosen according to the available DNS data at $Re_{\tau}=395$~\cite{Moser99}, $Re_{\tau}=950$~\cite{Alamo03}, $Re_{\tau}=2{,}000$~\cite{Hoyas06}, $Re_{\tau}=5{,}200$~\cite{Lee15}, and additionally $Re_{\tau}=10{,}000$, $Re_{\tau}=20{,}000$, and $Re_{\tau}=50{,}000$. All simulation cases, meshes, and resolution criteria are presented in Table~\ref{tab:ch_flows}.

The meshes considered are chosen such that the wall-parallel grid length scale yields approximately $\Delta=0.08\delta$ for most cases, so the RANS--LES switching point is located at $C_{\mathrm{DES}}\Delta=0.05\delta$. One simulation case uses twice the number of grid cells in streamwise and spanwise direction, resulting in a RANS--LES switching point near $C_{\mathrm{DES}}\Delta=0.025\delta$. As for the wall-normal resolution, the enrichment is taken into account in the wall-nearest cell layer in all simulation cases, see Figure~\ref{fig:ch_mesh}. As it was discussed earlier, the enrichment shape functions allow the resolution of the averaged near-wall flow with very coarse cell sizes. The width of the first off-wall cell lies in this work in the range of 51 to 191 wall units. In order to enable an application to high Reynolds numbers, a hyperbolic grid stretching is additionally considered, according to $f$: $[0,1] \to [-\delta, \delta]$:
\begin{equation}
x_2 \mapsto f(x_2)=\delta \frac{\tanh(\gamma (2x_2-1))}{\tanh(\gamma)},
\end{equation}
with the mesh stretching parameter $\gamma$. The values of $\gamma$ for all simulation cases are included in Table~\ref{tab:ch_flows}. In the numerical method, the velocity solution is postprocessed at a large number of wall-normal layers inside each cell using the definition of the velocity variable~\eqref{eq:dg_xw_rans:space} such that the behavior of the enrichment may be analyzed. Statistics were acquired in a simulation time interval of approximately 60--95 flow-through times based on a fixed time interval.

\begin{table*}[t]
\caption{Simulation cases and resolutions of the periodic hill flow. The cases use a coarse mesh with $32 {\times} 16 {\times} 16$ grid cells and a fine mesh with $64 {\times} 32 {\times} 32$ elements. The polynomial degree is $k=4$ for all simulation cases, and the number of grid points per direction is $k+1$ in each cell. The separation and reattachment lengths $x_{1,\mathrm{sep}}$ and $x_{1,\mathrm{reatt}}$ correspond to the zero-crossings of the skin friction.}
\label{tab:ph_flows}
\begin{tabular*}{\textwidth}{l @{\extracolsep{\fill}} l l l l l l l}
\hline
Case   & $N_{e1} {\times} N_{e2} {\times} N_{e3}$ &$N_{1} {\times} N_{2} {\times} N_{3}$ &$Re_{H}$ & $\mathrm{max}(\Delta y^+_{1e})$ & $x_{1,\mathrm{sep}}/H$ & $x_{1,\mathrm{reatt}}/H$
\\ \hline \noalign{\smallskip} 
ph10595\_coarse \ & $32 {\times} 16 {\times} 16$ & $160 {\times} 80 {\times} 80$ &  $10{,}595$ & $76$ & $0.25$ & $4.51$\\
ph10595\_fine \ & $64 {\times} 32 {\times} 32$ & $320 {\times} 160 {\times} 160$ &  $10{,}595$ & $36$  & $0.16$ & $4.40$\\
KKW\_DNS~\cite{Krank17b} \ & -  & $896 {\times} 448 {\times} 448$ & $10{,}595$ & -  & $0.19$ & $4.51$ \\
\noalign{\smallskip}  
 ph37000\_coarse \ & $32 {\times} 16 {\times} 16$ & $160 {\times} 80 {\times} 80$ &  $37{,}000$ & $144$  & $0.40$ & $3.37$\\
ph37000\_fine \ & $64 {\times} 32 {\times} 32$ & $320 {\times} 160 {\times} 160$ &  $37{,}000$ & $79$  & $0.26$ & $4.53$\\
 RM\_Exp~\cite{Rapp11} & - & - & $37{,}000$ & - & - & $3.76$\\
 CM\_WMLES\_coarse~\cite{Wiart17} & - & $128 {\times} 64 {\times} 64$ & $37{,}000$ & - & - & $2.3$\\
 CM\_WMLES\_fine~\cite{Wiart17} & - & $256 {\times} 128 {\times} 128$ & $37{,}000$ & - &  - & $2.8$\\ 
\hline
\end{tabular*}
\end{table*}

\begin{figure}[tb]
\centering
\includegraphics[trim= 10mm 40mm 10mm 40mm,clip,width=0.95\linewidth]{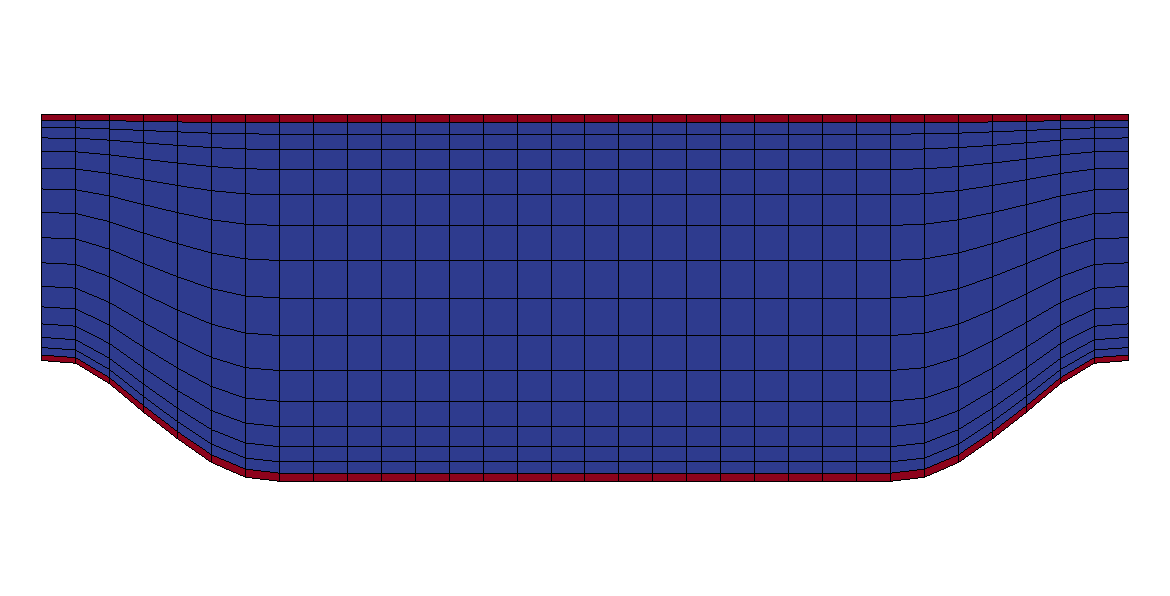}
\begin{picture}(100,0)
\put(-44,16){\footnotesize $x_1$}
\put(-62,32){\footnotesize $x_2$}
\end{picture}
\begin{tikzpicture}[overlay]
\draw[->,black, thick,>=latex]   
        (-5.87,0.44) -- (-5,0.44);
\draw[->,black, thick,rotate around={90:(-5.87,0.44)},>=latex]   
        (-5.87,0.44) -- (-5,0.44);
\end{tikzpicture}
\caption{Mesh for flow over periodic hills of the case ph37000\_coarse. Red indicates enriched cells and blue standard polynomial cells, i.e., a single layer of cells at the wall is enriched. In each cell, the solution consists of a polynomial of $4^{\mathrm{th}}$ degree plus one enrichment shape function in the enriched cells.}
\label{fig:ph_mesh}
\includegraphics[trim= 10mm 40mm 10mm 30mm,clip,width=0.95\linewidth]{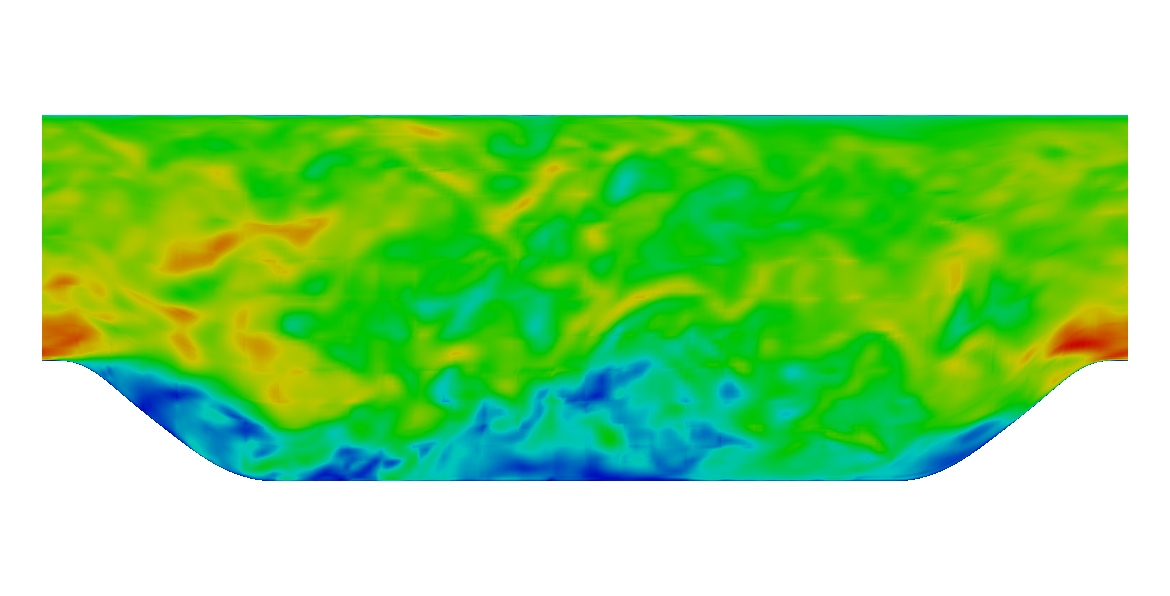}
\begin{picture}(100,0)
\put(-44,16){\footnotesize $x_1$}
\put(-62,32){\footnotesize $x_2$}
\end{picture}
\begin{tikzpicture}[overlay]
\draw[->,black, thick,>=latex]   
        (-5.87,0.44) -- (-5,0.44);
\draw[->,black, thick,rotate around={90:(-5.87,0.44)},>=latex]   
        (-5.87,0.44) -- (-5,0.44);
\end{tikzpicture}
\caption{Instantaneous numerical solution of flow over periodic hills of the case ph37000\_coarse via velocity magnitude. Red indicates high and blue low values.}
\label{fig:ph_flow}
\end{figure}

\begin{figure}[t]
\centering
\includegraphics[trim= 0mm 0mm 0mm 4mm,clip,width=0.7\linewidth]{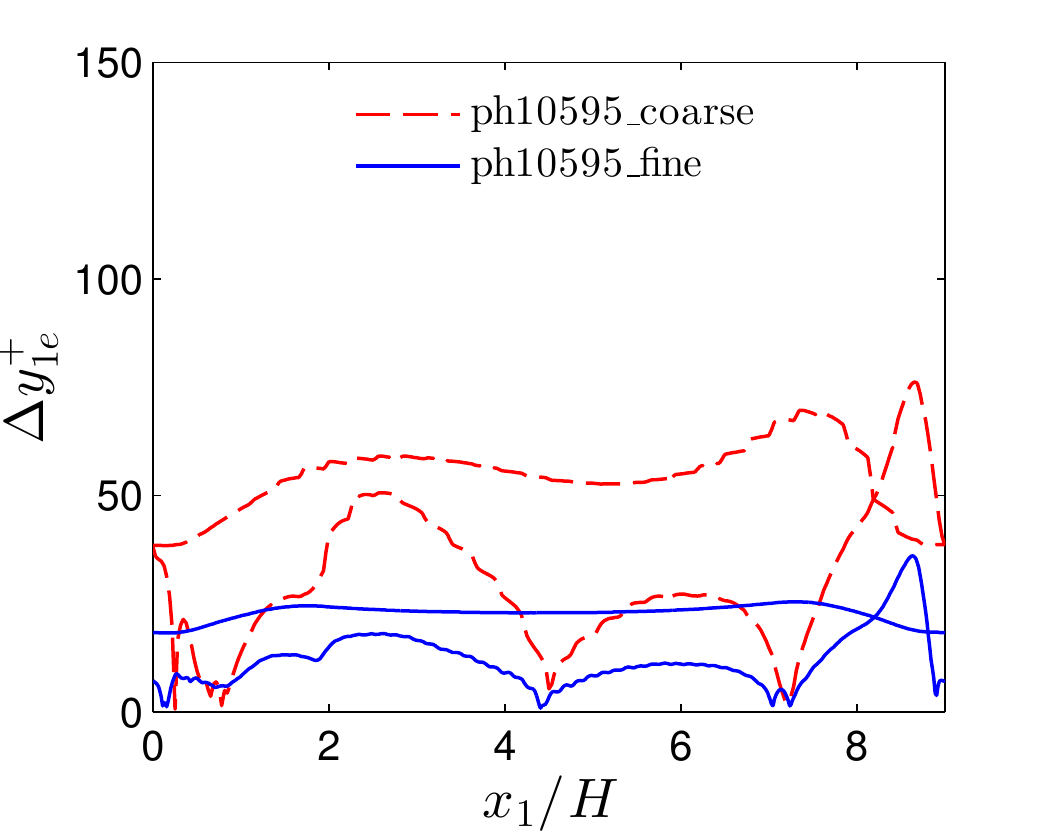}
\includegraphics[trim= 0mm 0mm 0mm 4mm,clip,width=0.7\linewidth]{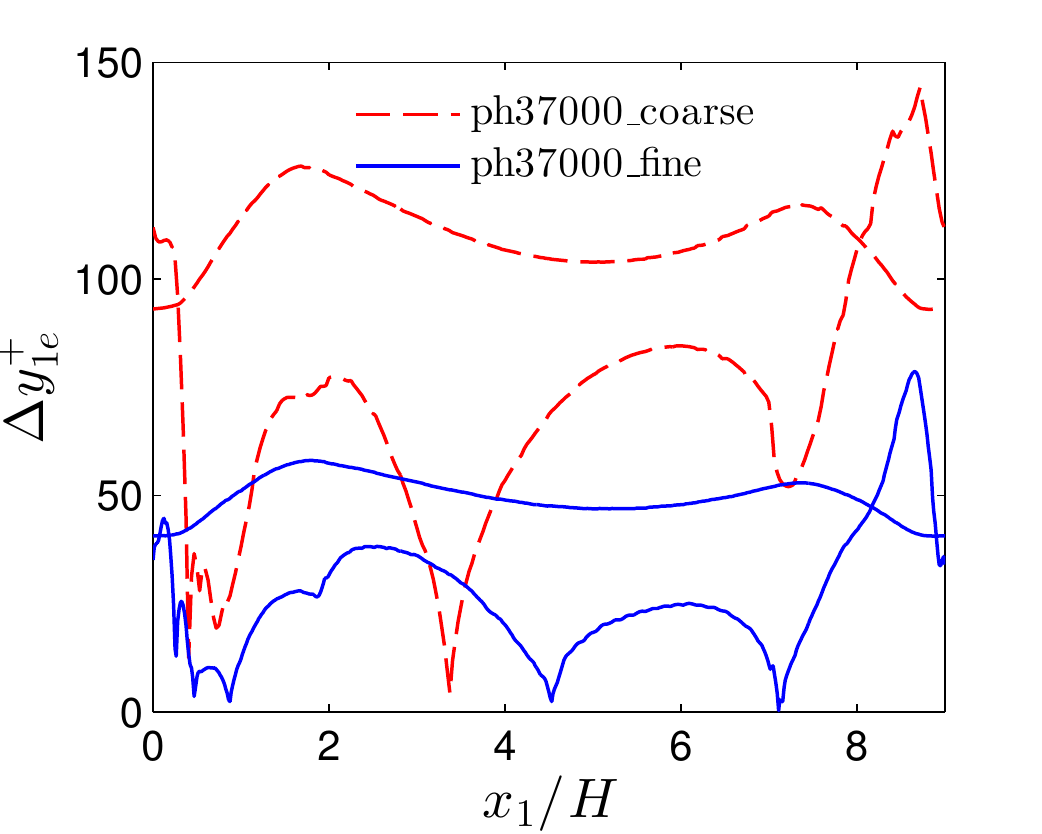}
\caption{Width of wall-layer (width of first off-wall cell) for $Re_H=10{,}595$ (top) and $Re_H=37{,}000$ (bottom). The shallower curves correspond to the upper wall.}
\label{fig:ph_yp1}
\end{figure}

\begin{figure*}[t!]
\centering
\includegraphics[trim= 9mm 0mm 9mm 4mm,clip,scale=0.65]{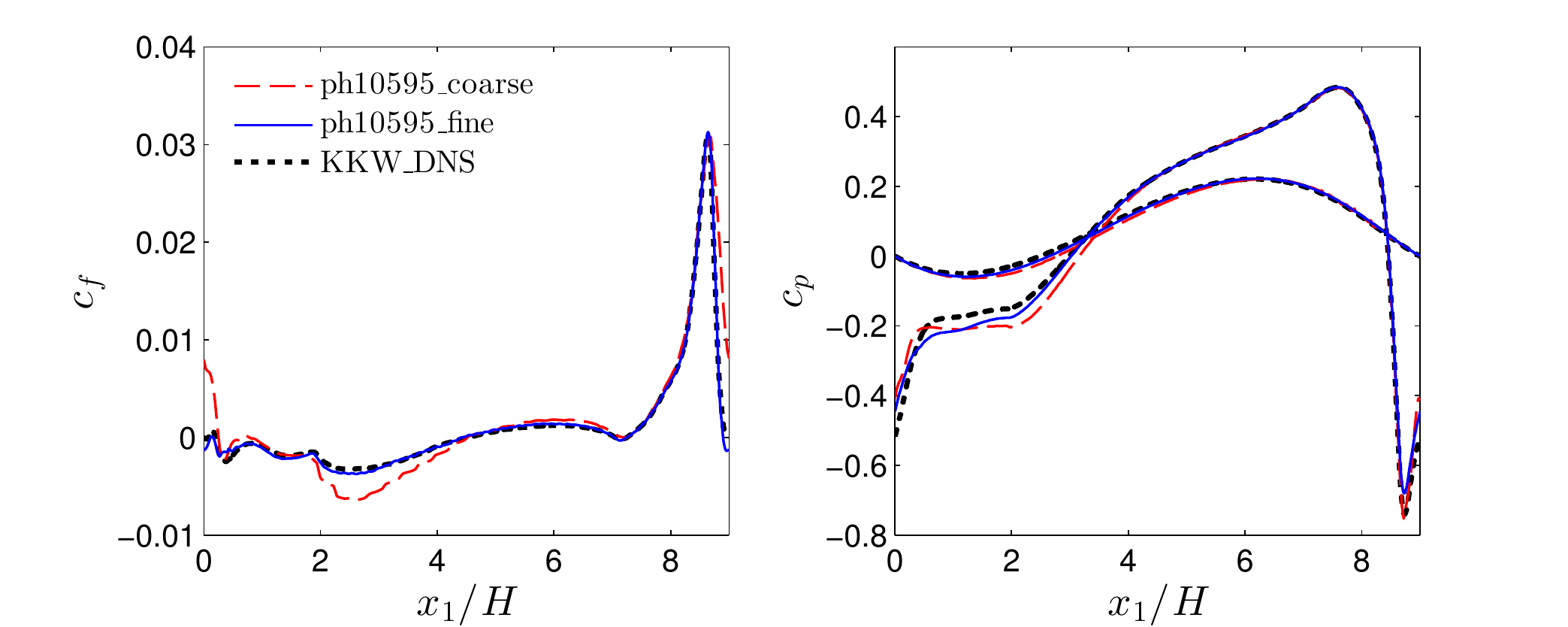}
\caption{Skin friction coefficient at the lower wall (left) and pressure coefficient at the lower and upper boundary (right). The shallower pressure coefficient curves correspond to the upper wall.}
\label{fig:ph10595_cfcp}
\end{figure*}

The turbulent flow is visualized at one time instant in Figure~\ref{fig:ch_flow}. Time-averaged results are presented in Figure~\ref{fig:ch_u}. Therein, the results are plotted in terms of the normalized mean velocity $u^+=\langle u_1\rangle/u_{\tau}$, the RMS velocity components $u^{\prime+}= \sqrt{\langle u_1^{\prime2}\rangle}/u_{\tau}$, $v^{\prime+}= \sqrt{\langle u_2^{2}\rangle}/u_{\tau}$, and $w^{\prime+}=\sqrt{\langle u_3^{2}\rangle}/u_{\tau}$, as well as the Reynolds shear stress $(u^{\prime}v^{\prime})^{+}=\langle u_1 u_2\rangle/u_{\tau}^2$, which are all normalized using the numerical value of $u_{\tau}$. The mean velocity is generally predicted very accurately in the laminar sublayer and the log-layer, where the enrichment shape functions are active. In order to get a better impression of the role of the enrichment, the numerical enrichment solution is plotted in Figure~\ref{fig:ch_u} alongside the full mean velocity solution. The enrichment solution represents the largest part of the near-wall solution in most cases, including the high velocity gradient. In particular in cases, where the first off-wall cell spans a range of more than 100 wall units, the enrichment is the main contributor to the mean velocity. Solely at the lowest Reynolds number, the enrichment solution plays a minor role, which essentially means that the polynomial component is capable of resolving most of the flow. Further away from the wall we observe the characteristic log-layer mismatch, that we expect in wall-attached simulations using DDES~\cite{Nikitin00,Yang17}. The log-layer mismatch is especially visible for the lower Reynolds numbers. We note that there are several techniques available in the literature that reduce this effect, for example~\cite{Shur08}. In the framework of the present enrichment methodology, it is possible to construct an alternative hybrid RANS/LES turbulence model, which does not show a log-layer mismatch by definition. We have recently developed such an approach, which is the topic of a subsequent publication~\cite{Krank17}.

The RMS velocities and the Reynolds shear stress are also presented in Figure~\ref{fig:ch_u} up to $Re_{\tau}=5{,}200$ and compared with the DNS data. These quantities show that the RANS--LES transition extends up to approximately $0.4\delta$ and the flow is in full LES mode further away from the wall. This means that we do not expect agreement with the DNS below $0.4\delta$, and the curves match the DNS above this value very well. Only in the refined case at $Re_{\tau}=2{,}000$, the RANS--LES transition happens closer to the wall.

Finally, a major advantage of the present method is the accurate prediction of the wall shear stress. In Table~\ref{tab:ch_flows}, we list the relative error of the computed wall shear stress compared to the nominal simulation parameters for each simulation case. The error lies within a few percent for all cases. Comparing the values with the errors in the skin friction coefficient presented in~\cite{Nikitin00} of up to 22\%, this is an excellent result.

We conclude from this section that wall modeling via function enrichment allows an accurate computation of the near-wall region in turbulent boundary layers with very coarse cells, while still computing the full incompressible Navier--Stokes equations in the whole boundary layer. DDES is a suitable turbulence modeling approach for wall modeling via function enrichment.

\begin{figure*}[t!]
\centering
\includegraphics[trim= 0mm 0mm 0mm 0mm,clip,width=0.65\linewidth]{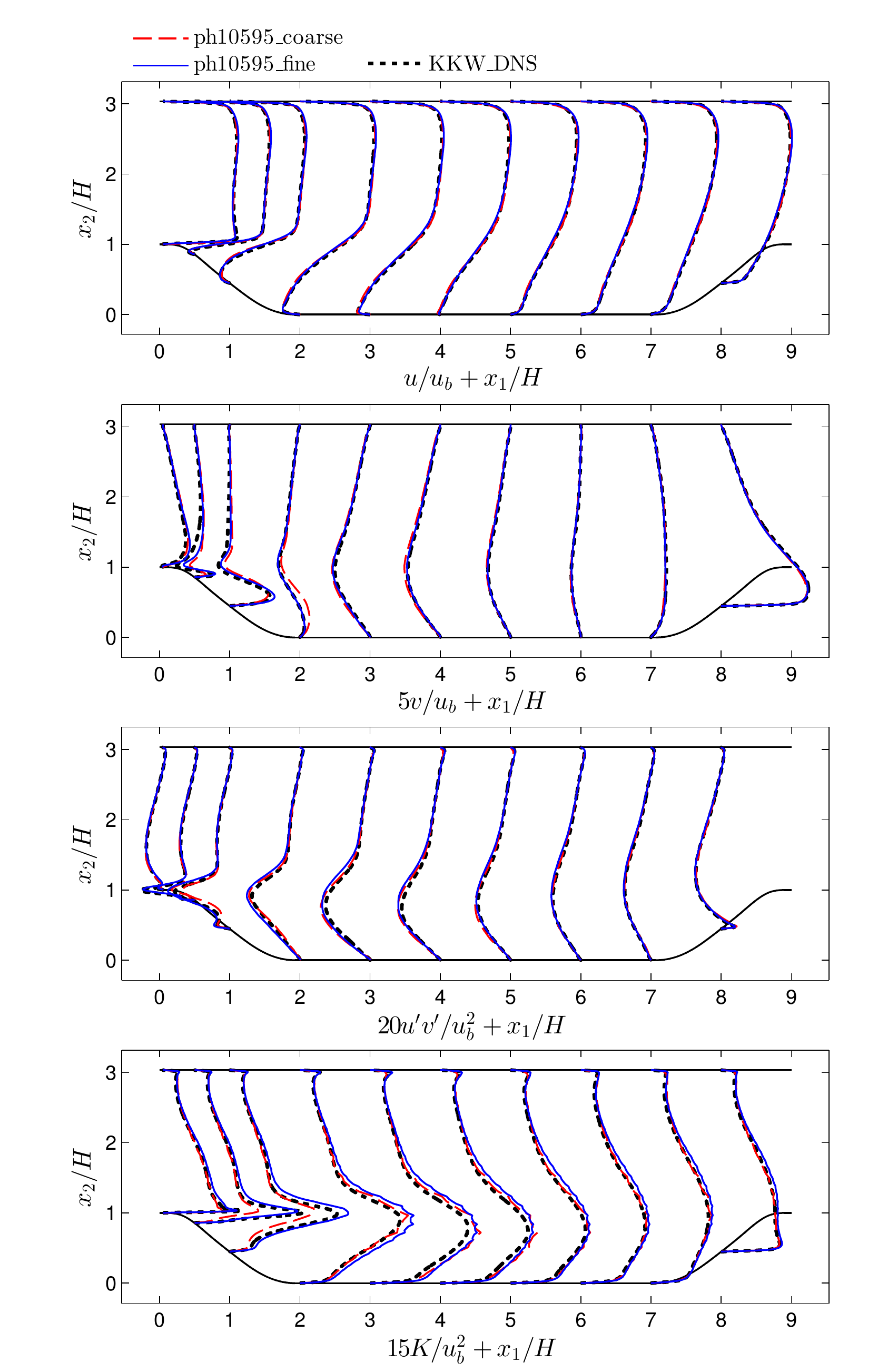}
\caption{Streamwise $u=\langle u_1\rangle$ and vertical $v=\langle u_2\rangle$ mean velocity, Reynolds shear stress $u'v'=\langle u_1 u_2\rangle - \langle u_1 \rangle \langle u_2\rangle$, and turbulence kinetic energy $K=1/2(u'u'+v'v'+w'w')$ of the periodic hill flow at $Re_H=10{,}595$.}
\label{fig:ph10595_um}
\end{figure*}

\begin{figure*}[t]
\centering
\includegraphics[trim= 0mm 63mm 0mm 0mm,clip,width=0.65\linewidth]{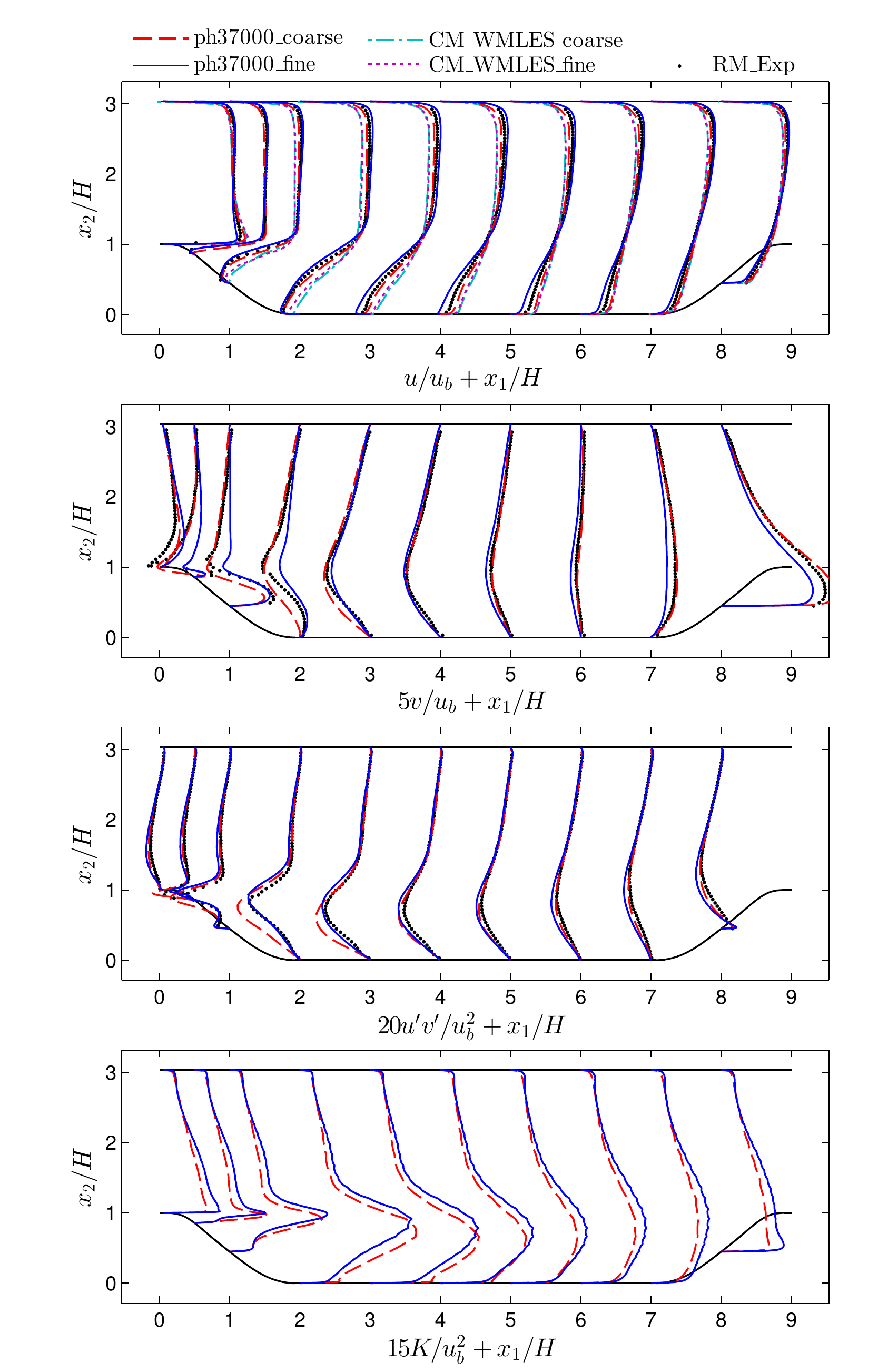}
\caption{Streamwise $u=\langle u_1\rangle$ and vertical $v=\langle u_2\rangle$ mean velocity as well as Reynolds shear stress $u'v'=\langle u_1 u_2\rangle - \langle u_1 \rangle \langle u_2\rangle$ of the periodic hill flow at $Re_H=37{,}000$. The results of the cases CM\_WMLES\_coarse and CM\_WMLES\_fine are only available for the streamwise velocity.}
\label{fig:ph37000_um}
\end{figure*}

\subsection{Flow over periodic hills}
As a second benchmark example, we consider flow over periodic hills at the Reynolds numbers based on the hill height $H$ and bulk velocity $u_b$ of $Re_H=10{,}595$ and $Re_H=37{,}000$. Several hybrid RANS/LES methods were assessed using this flow configuration within the European initiative ``Advanced Turbulence Simulation for Aerodynamic Application Challenges'' (ATAAC)~\cite{Schwamborn12}, including DDES (see the final report by Jakirli{\'c} for cross-comparison of results). A strong adverse pressure gradient and flow separation from the curved boundary are challenging for many statistical modeling approaches, but DDES yielded very good agreement with a reference LES in that study. Also, all previous publications on wall modeling via function enrichment~\cite{Krank16,Krank16c,Krank17} used this benchmark example, and very promising results were obtained if a turbulence resolving approach was used. Reference data for this flow is provided by DNS at the lower Reynolds number~\cite{Krank17b} (available for download at~\cite{Krank17d}) and water-channel experiments~\cite{Rapp11} at the higher Reynolds number.

The computational domain is of the dimensions $9H{\times}3.036H{\times}4.5H$ in streamwise, vertical and spanwise direction, respectively, and the lower wall is given by the smoothly curved hill shape. The domain is extended periodically in the streamwise and spanwise direction, and no-slip boundary conditions are applied on the upper and lower wall. The computational setup is very similar to the simulations of the DNS~\cite{Krank17b}. Two meshes are considered at each Reynolds number, a coarser mesh with $32{\times}16{\times}16$ cells, and a finer one with $64{\times}32{\times}32$ cells. As for the previous example, the solution is represented by a polynomial of degree 4 in each cell, plus one enrichment shape function in the wall-nearest cell layer. The mesh is moderately stretched towards the no slip walls to yield a better resolution of the near-wall area, and the geometry is mapped onto the exact hill shape using an isogeometric approach. One representative mesh is displayed in Figure~\ref{fig:ph_mesh}. The wall-normal width of the enrichment layer is plotted in Figure~\ref{fig:ph_yp1} in wall coordinates. An overview of all simulation cases and resolution parameters is given in Table~\ref{tab:ph_flows}. Statistics were averaged in a simulation time interval of 61 flow-through times. One snapshot of the instantaneous velocity field is visualized in Figure~\ref{fig:ph_flow}.

We begin the discussion of the results with the skin friction and pressure coefficients $c_f$ and $c_p$. They are defined as
\begin{equation*}
c_f=\frac{\tau_w}{\frac{1}{2} \rho u_b^2},\hspace{1cm}
c_p=\frac{p-p_{\mathrm{ref}}}{\frac{1}{2} \rho u_b^2},
\label{eq:cfcpdef}
\end{equation*}
where the reference pressure $p_{\mathrm{ref}}$ is taken at $x_1=0$ at the upper wall. The results of the lower Reynolds number are compared to the DNS in Figure~\ref{fig:ph10595_cfcp}. All profiles yield very good agreement with the DNS. Solely the skin friction coefficient predicted by the coarse mesh shows an overprediction of the magnitude between $x_1/H=2$ and $x_1/H=4$. Even the characteristic peak in the skin friction on the windward side of the hill crest is predicted very well for both cases. The overall excellent agreement is also observed in the estimation of the length of the reattachment zone of $x_{1,\mathrm{reatt}}/H=4.51$ and $4.40$ (see Table~\ref{tab:ph_flows}) in comparison to the DNS result of $x_{1,\mathrm{reatt}}/H=4.51\pm0.06$.

The velocity profiles of the same Reynolds number are compared to the DNS data at ten streamwise stations in Figure~\ref{fig:ph10595_um}. The streamwise velocity agrees exceptionally well with the reference DNS. The vertical velocity shows a minor difference at $x_1/H=2$ for the coarser simulation case, but the remaining profiles essentially lie on the DNS curves. A similar level of accuracy is observed in the Reynolds shear stress distribution. The turbulence kinetic energy computed with the coarse simulation case shows an underprediction of the magnitude in the shear layer. These results also exhibit ticks in the shear layer, which are typical for high-order DG, since the discontinuity present in the velocity yields higher fluctuations near the element boundaries, see also~\cite{Krank17b}.

The excellent results obtained at the lower Reynolds number motivate an application of the wall model to a significantly higher Reynolds number. The velocity statistics are compared to the available experimental reference data at $Re_H=37{,}000$ in Figure~\ref{fig:ph37000_um}. In order to allow for a critical assessment of the present wall modeling approach, we additionally compare the results of the mean streamwise velocity with a recent implementation of an equilibrium wall model within the high-order DG~\cite{Wiart17} (cases baseline and fine in that publication). These simulations employ grids comparable to the respective coarse and fine case presented in this work and are also included in the overview if Table~\ref{tab:ph_flows}. Regarding the mean velocity, all wall-modeled cases yield larger errors as compared to the lower Reynolds number. The equilibrium wall model overpredicts the velocity in the recirculation zone, yielding a shorter reattachment length of $x_{1,\mathrm{reatt}}/H=2.3$ and $x_{1,\mathrm{reatt}}/H=2.8$ in comparison to the experiments ($x_{1,\mathrm{reatt}}/H=3.76$, see Table~\ref{tab:ph_flows}). The present wall-enriched DDES simulations overpredict the mean streamwise velocity in that region with the coarse mesh and underpredict the velocity in the fine case. Yet, the DDES cases are closer to the reference than the equilibrium model, both for the coarse and fine mesh. The reattachment lengths are computed as $x_{1,\mathrm{reatt}}/H=3.37$ and $4.53$ and confirm the observations of the mean velocity. The profiles of the vertical velocity yield differences with the reference on the lee side of the hill as a result of the different length of the separation bubble. The magnitude of the Reynolds shear stress is overpredicted by the coarse case and is accurately estimated by the fine case.

We conclude from the results of the periodic hill flow that wall modeling via function enrichment with DDES as turbulence model is well capable of computing nonequilibrium flows. This is due to the full consistency of the method, as all terms of the Navier--Stokes equations are satisfied discretely.

\section{Conclusions}
In this work, we have used the DDES methodology to model the unresolved turbulent motions in wall modeling via function enrichment. The idea of this wall model is that an additional shape function is included in each cell, which has the shape of a wall function. As a result, the Galerkin method can resolve typical attached boundary layer profiles with very coarse meshes. Since the standard high-order polynomial shape functions are still available in all cells, the method is sufficiently flexible to represent nonequilibrium boundary layers with a high pressure gradient and separated boundary layers.

Wall modeling via function enrichment with the DDES turbulence model does not provide a solution to the problems in the hybrid RANS--LES transition region in attached boundary layers. However, an alternative hybrid RANS/LES turbulence modeling approach can be constructed based on the enrichment, which a priori circumvents these problems and the associated log-layer mismatch. This turbulence model is described in a follow-up paper~\cite{Krank17}.

\section*{Acknowledgements}
Computational resources on SuperMUC in Garching, Germany, provided by the Leibniz Supercomputing Centre, under the project pr83te are gratefully acknowledged. 

\end{document}